\documentclass[12pt]{article}
\usepackage{amsmath}
\usepackage{amssymb,scalefnt}
\usepackage{amsfonts}
\usepackage{latexsym}
\usepackage{color}
\usepackage{dsfont}
\usepackage{graphicx}

\catcode `\@=11 \@addtoreset{equation}{section}
\def\theequation{\arabic{section}.\arabic{equation}}
\catcode `\@=12
\newcommand{\be}{\begin{equation}}
\newcommand{\en}{\end{equation}}
\newcommand{\bea}{\begin{eqnarray}}
\newcommand{\ena}{\end{eqnarray}}
\newcommand{\beano}{\begin{eqnarray*}}
\newcommand{\enano}{\end{eqnarray*}}
\newcommand{\bee}{\begin{enumerate}}
\newcommand{\ene}{\end{enumerate}}

\newcommand{\Pc}{{\cal P}}
\newcommand{\Rc}{{\cal R}}

\catcode `\@=11 \@addtoreset{equation}{section}
\catcode `\@=12
\begin{document}

\begin{center}
{\Large \textbf{First results on applying a non-linear effect formalism to
alliances between political parties and buy and sell dynamics}}\vspace{1.5cm}%
\\[0pt]

{\large F. Bagarello}\\[0pt]
Dipartimento di Energia, Ingegneria dell'Informazione e Modelli Matematici,\\%
[0pt]
Scuola Politecnica Ingegneria, Universit\`a di Palermo,\\[0pt]
I-90128 Palermo, Italy\\[0pt]
and I.N.F.N., Sezione di Torino\\[0pt]
e-mail: fabio.bagarello@unipa.it\\[0pt]
home page: www.unipa.it/fabio.bagarello

\vspace{2mm}

{\large E. Haven}\\[0pt]
School of Management; Institute of Finance and IQSCS\\[0pt]
University of Leicester, LE1 7RH Leicester, UK\\[0pt]
e-mail: eh76@leicester.ac.uk\\[0pt]
\end{center}

\vspace*{.5cm}

\begin{abstract}
\noindent We discuss a non linear extension of a model of alliances in
politics, recently proposed by one of us. The model is constructed in terms
of operators, describing the \emph{interest} of three parties to form, or
not, some political alliance with the other parties. The time evolution of
what we call \emph{the decision functions} is deduced by introducing a
suitable hamiltonian, which describes the main effects of the interactions
of the parties amongst themselves and with their \emph{environments},
{which are }generated by their electors and by people who still have
no clear {idea }for which party to vote (or even if to vote). The
hamiltonian contains some non-linear effects, which takes into account the
role of a party in the decision process of the other two parties.

Moreover, we show how the same hamiltonian can also be used {to
construct a formal structure which can describe the dynamics of buying and
selling financial assets (without however implying a specific price setting
mechanism).}
\end{abstract}

\thispagestyle{empty}


\vspace{2cm}


\vfill


\newpage

\section{Introduction}

In a recent paper, \cite{all1}, a model of interaction between political
parties has been proposed. The model describes a decision making
procedure, deducing the time evolution of three so-called \emph{decision
functions} (DFs), one for each part{y} considered in our system.
These functions describe the interest of each party to form or not an
alliance with some other party. Their decisions are driven by the
interaction of each party with the other parties, with their own electors,
and with a set of undecided voters (i.e. people who have not yet decided to
vote for which party {(if at all they decide to vote))}. The approach
adopted in \cite{all1} uses an operatorial framework {(see also} \cite%
{bagbook}), in which the DFs are suitable mean values of certain number
operators associated to the parties. The dynamics {are} driven by a
suitable hamiltonian which implements the various interactions between the
different actors of the system.

The limitation of the model, as described in \cite{all1}, is that the
hamiltonian is quadratic and, as a consequence, the equations of motion are
linear. This simplifies quite a bit the analysis of the time evolution of
the system. {I}n fact an exact solution can be deduced in that case,
but the price we pay is that the model is not entirely realistic, since the
hamiltonian does not include contributions which might be relevant in a
concrete situation. In this paper we introduce several \textit{non-linear
contributions} in the model, and we solve, adopting a suitable
approximation, the related non-linear differential equations. These
non-linear terms are needed to introduce in the model some sort of three-bod%
{y} interactions, which were not included in \cite{all1}. The reason
why these terms are interesting is because they describe (please see below
for more details) the role of, say, the first party ($\mathcal{P}_{1}$), in
the explicit strength of the interaction between the other two parties, $%
\mathcal{P}_{2}$ and $\mathcal{P}_{3}$. This is important, since it is
natural to assume that the DFs of both $\mathcal{P}_{2}$ and $\mathcal{P}_{3}
$ also depend on what $\mathcal{P}_{1}$ is doing.

It is important to notice that not many contributions exist in the
mathematical and physic{s} literature on politics, and only {%
very }few of them adopt a quantum mechanical (or operator) point of view, as
the one used in \cite{all1}. We refer to \cite%
{otto,havkhre,galam1,galam2} for some recent and not so recent
contributions on this topic.

After a long discussion on politics, we also show how the same hamiltonian
can be used, with just some minor changes, to deduce the dynamics of a
buy-and-sell financial system.

The paper is organized as follows: in the next section we introduce the
model, we derive the differential equations and we propose an approximation
scheme to solve them. In Section III we show how to model a {simple }%
financial system using the same general settings. Section IV contains our
conclusions. To keep the paper self-contained, and to make it {also }%
more readable to those who are not familiar with quantum mechanics, we have
added an appendix where a few crucial aspects of operators and quantum
dynamics are reviewed.

\section{Modelling alliances in politics and its dynamics}

\label{sectII}

In this section we discuss the details of our model {and we will
first }construct the vectors describing the players and the hamiltonian of
the system. {We then }deduce the differential equations of motion. To
keep the paper self contained, we recall first {a }few important
facts which were already discussed in \cite{all1}.

In our system we have three parties, $\mathcal{P}_{1}$, $\mathcal{P}_{2}$
and $\mathcal{P}_{3}$, which, together, form the system $\mathcal{S}_{%
\mathcal{P}}$. Each party has to make a choice, and it can choose only `one'
or `zero', {which }corresponds respectively to {either }\emph{%
form a coalition} or not. This is, in fact, the only aspect of the parties
we are interested in. Hence, we have eight different possibilities, {%
to }which we associate eight different and mutually orthogonal vectors in a%
{n} eight-dimensional Hilbert space $\mathcal{H}_{\mathcal{P}}$.
These vectors are $\varphi _{i,k,l}$, with $i,k,l=0,1$. {As an
example, }the first vector, $\varphi _{0,0,0}$, describes the fact that, at $%
t=0$, no party wants to ally with the other parties. Of course, this
attitude can change during the time evolution. {What is interesting
to know is: }how does this {attitude }change? And how {can one}
describe this change? {Let us consider another example. F}or
instance, $\varphi _{0,1,0}$, describes the fact that, at $t=0$, $\mathcal{P}%
_{1}$ and $\mathcal{P}_{3}$ do {not want} to form any coalition,
while $\mathcal{P}_{2}$ does. $\mathcal{F}_{\varphi }=\{\varphi
_{i,k,l},\,i,k,l=0,1\}$ is an orthonormal basis for $\mathcal{H}_{\mathcal{P}%
}$. A generic vector of $\mathcal{S}_{\mathcal{P}}$, for $t=0$, is a linear
combination of the form
\begin{equation}
\Psi =\sum_{i,k,l=0}^{1}\alpha _{i,k,l}\varphi _{i,k,l},  \label{24}
\end{equation}%
where we assume $\sum_{i,k,l=0}^{1}|\alpha _{i,k,l}|^{2}=1$ in order to
normalize the total probability, \cite{khren2}. In particular, for instance,
$|\alpha _{0,0,0}|^{2}$ represents the probability that $\mathcal{S}_{%
\mathcal{P}}$ is, at $t=0$, in a state $\varphi _{0,0,0}$, i.e. that $%
\mathcal{P}_{1}$, $\mathcal{P}_{2}$ and $\mathcal{P}_{3}$ have chosen `0'
(no coalition).

As in \cite{all1}, and for the same reasons (see below), we construct the
vectors $\varphi _{i,k,l}$ in a very special way, starting with the vacuum
of three fermionic operators, $p_{1}$, $p_{2}$ and $p_{3}$, i.e. three
operators which, together with their adjoint, satisfy the canonical
anticommutation relation (CAR) $\{p_{k},p_{l}^{\dagger }\}=\delta _{k,l}$
and $\{p_{k},p_{l}\}=0$. Here $\{x,y\}=xy+yx$, for all pairs $x$ and $y$.
More in detail, $\varphi _{0,0,0}$ is such that $p_{j}\varphi _{0,0,0}=0$, $%
j=1,2,3$. The other vectors $\varphi _{i,j,k}$ can be constructed acting on $%
\varphi _{0,0,0}$ with the operators $p_{1}^{\dagger }$, $p_{2}^{\dagger }$
and $p_{3}^{\dagger }$:
\begin{equation*}
\varphi _{1,0,0}=p_{1}^{\dagger }\varphi _{0,0,0},\quad \varphi
_{0,1,0}=p_{2}^{\dagger }\varphi _{0,0,0},\quad \varphi
_{1,1,0}=p_{1}^{\dagger }\,p_{2}^{\dagger }\varphi _{0,0,0},\quad \varphi
_{1,1,1}=p_{1}^{\dagger }\,p_{2}^{\dagger }\,p_{3}^{\dagger }\varphi
_{0,0,0},
\end{equation*}%
and so on. Let now $\hat{P}_{j}=p_{j}^{\dagger }p_{j}$ be the so-called
\emph{number operator} of the $j$-th party, which is constructed using $p_{j}
$ and its adjoint, $p_{j}^{\dagger }$. Since $\hat{P}_{j}\varphi
_{n_{1},n_{2},n_{3}}=n_{j}\varphi _{n_{1},n_{2},n_{3}}$, for $j=1,2,3$, it
is clear that $\varphi _{n_{1},n_{2},n_{3}}$ are eigenvectors of these
operators, while their eigenvalues, zero and one, correspond to the only
possible choices admitted for the three parties at $t=0$. This is, in fact,
the main reason why we have used here the fermionic operators $p_{j}$: they
automatically produce only these eigenvalues. Our first effort now consists
in \emph{giving a dynamics} to the number operators $\hat{P}_{j}$, following
the general scheme proposed in \cite{bagbook}. Hence, we look for an
Hamiltonian $H$ which describes the interactions between the various
constituents of the system. Once $H$ is given, we can compute first the time
evolution of the number operators as $\hat{P}_{j}(t):=e^{iHt}\hat{P}%
_{j}e^{-iHt}$, and {we can }then {ascertain }their mean values
on some suitable state describing the system at $t=0$, in order to get what
we have already called \emph{decision functions}, (DFs) ({please }see
below). The \emph{rules} needed to write down $H$ are described in \cite%
{bagbook}, and adopted in \cite{all1} where it is also discussed why the
three parties are just part of a larger system which must {also }%
include the set of electors. In fact, it is mainly this interaction which
creates the final decision. Hence, $\mathcal{S}_{\mathcal{P}}$ must be \emph{%
open}, {and we mean }with this that there must {exist} some
\emph{large} environment, $\mathcal{R}$, {which }interact{s}
with $\mathcal{P}_{1}$, $\mathcal{P}_{2}$ and $\mathcal{P}_{3}$, {and
it} produces some sort of feedback used by $\mathcal{P}_{j}$ to decide.
Fermionic operators (depending also on a continuous index) are also used to
describe their environment, \cite{all1}.

\begin{figure}[tbp]
\begin{center}
\begin{picture}(450,90)

\put(160,65){\thicklines\line(1,0){45}}
\put(160,85){\thicklines\line(1,0){45}}
\put(160,65){\thicklines\line(0,1){20}}
\put(205,65){\thicklines\line(0,1){20}}
\put(183,75){\makebox(0,0){$\Pc_2$}}

\put(300,35){\thicklines\line(1,0){45}}
\put(300,55){\thicklines\line(1,0){45}}
\put(300,35){\thicklines\line(0,1){20}}
\put(345,35){\thicklines\line(0,1){20}}
\put(323,45){\makebox(0,0){$\Pc_3$}}

\put(10,35){\thicklines\line(1,0){45}}
\put(10,55){\thicklines\line(1,0){45}}
\put(10,35){\thicklines\line(0,1){20}}
\put(55,35){\thicklines\line(0,1){20}}
\put(33,45){\makebox(0,0){$\Pc_1$}}

\put(10,-55){\thicklines\line(1,0){45}}
\put(10,-35){\thicklines\line(1,0){45}}
\put(10,-55){\thicklines\line(0,1){20}}
\put(55,-55){\thicklines\line(0,1){20}}
\put(33,-45){\makebox(0,0){$\Rc_1$}}

\put(140,-55){\thicklines\line(1,0){85}}
\put(140,-35){\thicklines\line(1,0){85}}
\put(140,-55){\thicklines\line(0,1){20}}
\put(225,-55){\thicklines\line(0,1){20}}
\put(183,-45){\makebox(0,0){$\Rc_2$}}

\put(300,-55){\thicklines\line(1,0){45}}
\put(300,-35){\thicklines\line(1,0){45}}
\put(300,-55){\thicklines\line(0,1){20}}
\put(345,-55){\thicklines\line(0,1){20}}
\put(323,-45){\makebox(0,0){$\Rc_3$}}

\put(140,-155){\thicklines\line(1,0){85}}
\put(140,-95){\thicklines\line(1,0){85}}
\put(140,-155){\thicklines\line(0,1){60}}
\put(225,-155){\thicklines\line(0,1){60}}
\put(183,-125){\makebox(0,0){$\Rc_{und}$}}

\put(70,44){\thicklines\vector(1,0){220}}
\put(70,44){\thicklines\vector(-1,0){3}}
\put(70,44){\thicklines\vector(3,1){80}}
\put(70,44){\thicklines\vector(-3,-1){3}}
\put(290,44){\thicklines\vector(-3,1){80}}
\put(290,44){\thicklines\vector(3,-1){3}}

\put(31,27){\thicklines\vector(0,-1){55}}
\put(31,27){\thicklines\vector(0,1){3}}
\put(322,27){\thicklines\vector(0,-1){55}}
\put(322,27){\thicklines\vector(0,1){3}}
\put(165,57){\thicklines\vector(0,-1){85}}
\put(165,57){\thicklines\vector(0,1){3}}

\put(35,27){\thicklines\vector(1,-1){115}}
\put(35,27){\thicklines\vector(-1,1){3}}
\put(318,27){\thicklines\vector(-1,-1){115}}
\put(318,27){\thicklines\vector(1,1){3}}
\put(195,57){\thicklines\vector(0,-1){145}}
\put(195,57){\thicklines\vector(0,1){3}}


\end{picture}
\end{center}
\par
\vspace*{5.3cm}
\caption{The system and its multi-component reservoir.}
\label{figscheme}
\end{figure}
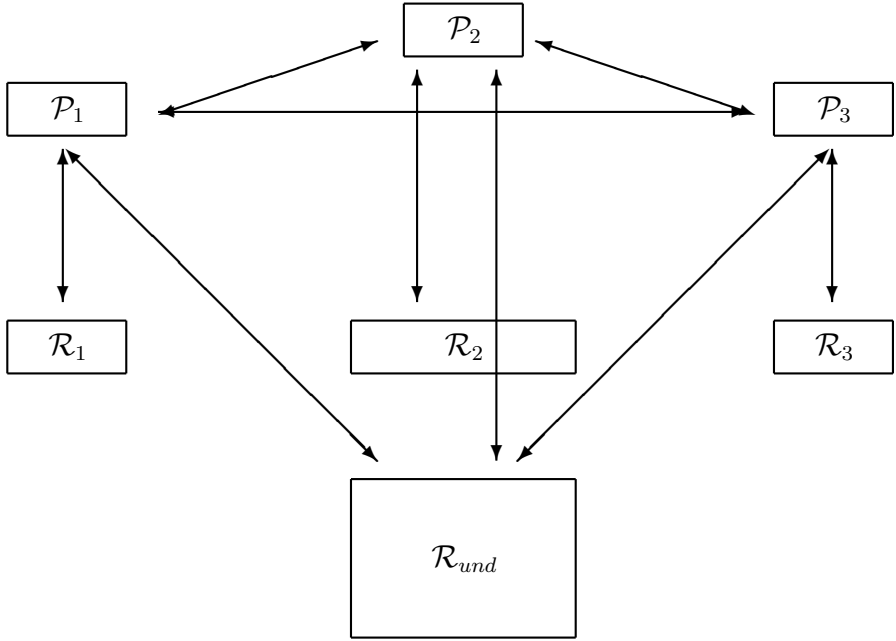

The various elements of our model are described in Figure \ref{figscheme},
where the various arrows show all the admissible interactions.

\vspace*{1cm}

In this figure $\mathcal{R}_{j}$ represents the set of the electors of $%
\mathcal{P}_{j}$, while $\mathcal{R}_{und}$ is the set of all the undecided
voters. Figure \ref{figscheme} shows, for instance, that $\mathcal{P}_{1}$
can interact with $\mathcal{R}_{1}$ and $\mathcal{R}_{und}$, but n{%
either} with $\mathcal{R}_{2}$ {n}or with $\mathcal{R}_{3}$. We also
see that $\mathcal{P}_{1}$ interacts with both $\mathcal{P}_{2}$ and $%
\mathcal{P}_{3}$. To define the hamiltonian which describes, in our
framework, the scheme in Figure \ref{figscheme}, we start introducing the
following purely quadratic operator, which is, essentially, the one adopted
in \cite{all1}:
\begin{equation}
\left\{
\begin{array}{ll}
h=H_{0}+H_{PBs}+H_{PB}+H_{int}, &  \\
H_{0}=\sum_{j=1}^{3}\omega _{j}p_{j}^{\dagger }p_{j}+\sum_{j=1}^{3}\int_{%
\mathbb{R}}\Omega _{j}(k)B_{j}^{\dagger }(k)B_{j}(k)\,dk+\int_{\mathbb{R}%
}\Omega (k)B^{\dagger }(k)B(k)\,dk, &  \\
H_{PBs}=\sum_{j=1}^{3}\lambda _{j}\int_{\mathbb{R}}\left(
p_{j}B_{j}^{\dagger }(k)+B_{j}(k)p_{j}^{\dagger }\right) \,dk, &  \\
H_{PB}=\sum_{j=1}^{3}\tilde{\lambda}_{j}\int_{\mathbb{R}}\left(
p_{j}B^{\dagger }(k)+B(k)p_{j}^{\dagger }\right) \,dk, &  \\
H_{int,l}=\mu _{12}^{(0)}\left( p_{1}^{\dagger }p_{2}+p_{2}^{\dagger
}p_{1}\right) +\nu _{12}^{(0)}\left( p_{1}^{\dagger }p_{2}^{\dagger
}+p_{2}p_{1}\right) +\mu _{13}^{(0)}\left( p_{1}^{\dagger
}p_{3}+p_{3}^{\dagger }p_{1}\right) + &  \\
\qquad +\nu _{13}^{(0)}\left( p_{1}^{\dagger }p_{3}^{\dagger
}+p_{3}p_{1}\right) +\mu _{23}^{(0)}\left( p_{2}^{\dagger
}p_{3}+p_{3}^{\dagger }p_{2}\right) +\nu _{23}^{(0)}\left( p_{2}^{\dagger
}p_{3}^{\dagger }+p_{3}p_{2}\right) . &
\end{array}%
\right.   \label{22}
\end{equation}%
Here $\omega _{j}$, $\lambda _{j}$, $\tilde{\lambda}_{j}$, $\mu _{ij}^{(0)}$
and $\nu _{ij}^{(0)}$ are real quantities, while $\Omega _{j}(k)$ and $%
\Omega (k)$ are real-valued functions. Their meaning is explained in detail
in \cite{all1}. As already anticipated, the following CAR's for the
operators of the reservoir are assumed:
\begin{equation}
\{B_{i}(k),B_{l}^{\dagger }(q)\}=\delta _{i,l}\delta (k-q)\,1\!\!1,\qquad
\{B_{i}(k),B_{l}(k)\}=0,  \label{23}
\end{equation}%
as well as
\begin{equation}
\{B(k),B^{\dagger }(q)\}=\delta (k-q)\,1\!\!1,\quad \{B(k),B(k)\}=0.
\label{23b}
\end{equation}%
Moreover each $p_{j}^{\sharp }$ anti-commutes with each $B_{j}^{\sharp }(k)$
and with $B^{\sharp }(k)$: $\{b_{j}^{\sharp },B_{l}^{\sharp
}(k)\}=\{b_{j}^{\sharp },B^{\sharp }(k)\}=0$ for all $j$, $l$ and for all $k$%
, and we further assume that $\{B^{\sharp }(q),B_{l}^{\sharp }(k)\}=0$. Here
$X^{\sharp }$ stands for $X$ or $X^{\dagger }$.

The full hamiltonian is now obtained by adding to $h$ another term, $\delta h
$, which contains some non quadratic terms:
\begin{equation}
\left\{
\begin{array}{ll}
H=h+\delta h, &  \\
\delta h=h_{int}^{ex}+h_{int}^{coop}, &  \\
h_{int}^{ex}=\left( \mu _{12}^{(2)}+(\mu _{12}^{(1)}-\mu
_{12}^{(2)})N_{3}\right) \left( p_{1}^{\dagger }p_{2}+p_{2}^{\dagger
}p_{1}\right) +\left( \mu _{13}^{(2)}+(\mu _{13}^{(1)}-\mu
_{13}^{(2)})N_{2}\right) \left( p_{1}^{\dagger }p_{3}+p_{3}^{\dagger
}p_{1}\right) + &  \\
\qquad +\left( \mu _{23}^{(2)}+(\mu _{23}^{(1)}-\mu _{23}^{(2)})N_{1}\right)
\left( p_{2}^{\dagger }p_{3}+p_{3}^{\dagger }p_{2}\right) , &  \\
h_{int}^{coop}=\left( \nu _{12}^{(2)}+(\nu _{12}^{(1)}-\nu
_{12}^{(2)})N_{3}\right) \left( p_{1}^{\dagger }p_{2}^{\dagger
}+p_{2}p_{1}\right) +\left( \nu _{13}^{(2)}+(\nu _{13}^{(1)}-\nu
_{13}^{(2)})N_{2}\right) \left( p_{1}^{\dagger }p_{3}^{\dagger
}+p_{3}p_{1}\right) + &  \\
\qquad +\left( \nu _{23}^{(2)}+(\nu _{23}^{(1)}-\nu _{23}^{(2)})N_{1}\right)
\left( p_{2}^{\dagger }p_{3}^{\dagger }+p_{3}p_{2}\right) , &
\end{array}%
\right.   \label{24b}
\end{equation}%
where, again, $\mu _{ij}^{(1)}$, $\mu _{ij}^{(2)}$, $\nu _{ij}^{(1)}$ and $%
\nu _{ij}^{(2)}$ are real quantities. Let us now explain the various terms
in $H$.

The first contribution in (\ref{22}) is $H_{0}$, which describes the free
evolution of the operators of $\mathcal{S}=\mathcal{S}_{\mathcal{P}}\otimes
\mathcal{R}$, where $\mathcal{R}=(\mathcal{R}_{1}\otimes \mathcal{R}%
_{2}\otimes \mathcal{R}_{3})\otimes \mathcal{R}_{und}$. If, in particular,
all the interaction parameters $\lambda _{j},\tilde{\lambda}_{j}$, $\mu
_{ij}^{(l)}$ and $\nu _{ij}^{(l)}$ are zero, then $H=H_{0}$. Hence, since in
this case $[H,\hat{P}_{j}]=0$, the number operators describing the choices
of the three parties (and their related DFs) stay constant in time. In other
words, in {the }absence of interactions, the original choice of each $%
\mathcal{P}_{j}$ is not affected by the time evolution. Translating this in
the Schr\"{o}dinger representation, this means that if $\mathcal{S}_{%
\mathcal{P}}$ is in an eigenstate $\varphi _{n_{1},n_{2},n_{3}}$ of $H_{0}$,
then it remains in the same state also for $t>0$. However, we should also
add that if $\mathcal{S}_{\mathcal{P}}$ is in the state $\Psi $ in (\ref{24}%
), we might have non trivial dynamics already at this level. As discussed in
\cite{all1}, $H_{PBs}$ describes the interaction between the three parties
and their related groups of electors: $p_{j}B_{j}^{\dagger }(k)$ describes
the fact that, when some sort of \emph{global reaction against alliance}
(GRAA) increases, then $\mathcal{P}_{j}$ tends to chose `0' (no coalition).
On the other hand, $B_{j}(k)p_{j}^{\dagger }$ describes the fact that $%
\mathcal{P}_{j}$ look{s} for some coalition when the GRAA of its
electors decreases. This is because of the raising and lowering operators $%
p_{j}^{\dagger }$ and $p_{j}$ in these interaction terms, coupled
respectively with the lowering ($B_{j}(k)$) and raising ($B_{j}^{\dagger }(k)
$) operators of the electors of $\mathcal{P}_{j}$. A similar interpretation
holds for $H_{PB}$, with the difference that the interaction is now between
the parties and a single set of undecided voters. The last contribution in $h
$, $H_{int,l}$, is introduced to describe the fact that the parties {%
also} {attempt} to talk to each other to get some agreement. Two
possibilities are allowed; \textbf{i)} the parties act \emph{cooperatively}
(they make the same choice, and we have terms like $p_{j}^{\dagger
}p_{k}^{\dagger }$), and; \textbf{ii) }they make opposite choices. For
instance $\mathcal{P}_{1}$ tr{ies} to form some alliance, while $%
\mathcal{P}_{2}$ excludes this possibility (and we have terms like $%
p_{1}^{\dagger }p_{2}$). Of course, the relative magnitude of $\mu
_{jk}^{(0)}$ and $\nu _{jk}^{(0)}$ decides which is the leading contribution
in $H_{int,l}$. It is important to stress that all the terms in $H_{int,l}$
are quadratic, so that the contributions they produce in the differential
Heisenberg equations turn out to be linear. This is the reason why it was
possible, in \cite{all1}, to produce an analytical solution for the time
evolution of the system. However, the extra terms in (\ref{24b}) make, in
our opinion, the situation more interesting from the point of view of the
real interpretation. In fact, whil{st} in $H_{int,l}$ the will of $%
\mathcal{P}_{1}$ to form or not an alliance with $\mathcal{P}_{2}$ is
totally independent of what $\mathcal{P}_{3}$ is doing, this is not so when
we also consider $\delta h$. For instance, let us {consider} the
interaction between $\mathcal{P}_{1}$ and $\mathcal{P}_{2}$, and in
particular {let us focus on} the \emph{exchange} term, which we now
rewrite as follows:%
\begin{equation*}
\left( \mu _{12}^{(2)}+(\mu _{12}^{(1)}-\mu _{12}^{(2)})N_{3}\right) \left(
p_{1}^{\dagger }p_{2}+p_{2}^{\dagger }p_{1}\right) =\mu
_{12}^{(1)}N_{3}\left( p_{1}^{\dagger }p_{2}+p_{2}^{\dagger }p_{1}\right)
+\mu _{12}^{(2)}(1\!\!1-N_{3})\left( p_{1}^{\dagger }p_{2}+p_{2}^{\dagger
}p_{1}\right) .
\end{equation*}%
The meaning of the two contributions is now evident: the first term, i.e.
the one proportional to $\mu _{12}^{(1)}$ in the RHS, describes the fact
that the more $\mathcal{P}_{3}$ is willing to ally with $\mathcal{P}_{1}$ or
$\mathcal{P}_{2}$, the more these two parties tend to behave differently:
one is \emph{pleased} with $\mathcal{P}_{3}$'s attentions, the other is not.
The other term, the one proportional to $\mu _{12}^{(2)}$, describes a
specula{tive} behavior. $\mathcal{P}_{1}$ and $\mathcal{P}_{2}$ tend
to behave differently when the interest of $\mathcal{P}_{3}$ to form a
coalition is low. In other words, what decides the relative strength of the $%
\mathcal{P}_{1}\leftrightarrow \mathcal{P}_{2}$ interaction is not (only)
the relative value of $\mu _{12}^{(1)}$ and $\mu _{12}^{(2)}$, but also, and
more interestingly, the attitude of $\mathcal{P}_{3}$ to form (or not) a
coalition. The behavior of $\mathcal{P}_{1}$ and $\mathcal{P}_{2}$ is
related also to what $\mathcal{P}_{3}$ is doing. Of course, a similar
analysis can be repeated for the other terms in $h_{int}^{ex}$, while for
what concerns $h_{int}^{coop}$ the presence of $N_{j}$ or $1\!\!1-N_{j}$
introduces, again, different weights in the various terms of the
hamiltonian. {However, }the other two parties {now }tend to
behave in the same way. For instance, rewriting
\begin{equation*}
\left( \nu _{12}^{(2)}+(\nu _{12}^{(1)}-\nu _{12}^{(2)})N_{3}\right) \left(
p_{1}^{\dagger }p_{2}^{\dagger }+p_{2}p_{1}\right) =\nu
_{12}^{(1)}N_{3}\left( p_{1}^{\dagger }p_{2}^{\dagger }+p_{2}p_{1}\right)
+\nu _{12}^{(2)}(1\!\!1-N_{3})\left( p_{1}^{\dagger }p_{2}^{\dagger
}+p_{2}p_{1}\right) ,
\end{equation*}%
we see that when $\mathcal{P}_{3}$ wants to form some coalition, then both $%
\mathcal{P}_{1}$ and $\mathcal{P}_{2}$ react in the same way. They both try
to form (or not to form) a coalition, with $\mathcal{P}_{3}$, or between
themselves. Moreover, we are also considering the possibility in which the
strength of the interaction is proportional to $1\!\!1-N_{3}$ rather than to
$N_{3}$. Of course, we stress again that other than the value of $N_{3}$,
what is also crucial in deciding the strength of the various terms in $%
\delta h$, {are} the numerical values of the parameters $\mu
_{ij}^{(k)}$ and $\nu _{ij}^{(k)}$.

\vspace{1mm}

We are now ready to continue with the analysis of the dynamics of the
system. The Heisenberg equations of motion $\dot X(t)=i[H,X(t)]$, \cite%
{bagbook}, can be deduced by using the CAR (\ref{23}) and (\ref{23b}) above.
The result can be written as follows:

\begin{equation}
\left\{
\begin{array}{ll}
\dot p_1(t)=l_1(t)+nl_1(t), &  \\
\vspace{1mm} \dot p_2(t)=l_2(t)+nl_2(t), &  \\
\vspace{1mm} \dot p_3(t)=l_3(t)+nl_3(t), &  \\
\vspace{1mm} \dot B_j(q,t)=-i\Omega_j(q) B_j(q,t)+i\lambda_j p_j(t),\qquad
j=1,2,3, &  \\
\vspace{1mm} \dot B(q,t)=-i\Omega(q) B(q,t)+i\sum_{j=1}^3\tilde\lambda_j
p_j(t), \label{26} &
\end{array}%
\right.
\end{equation}
where we have introduced the following quantities:
\begin{equation}
\left\{
\begin{array}{ll}
l_1(t)=-i\omega_1 p_1(t)+i\lambda_1\int_{\mathbb{R}}B_1(q,t)\,dq+i\tilde%
\lambda_1\int_{\mathbb{R}}B(q,t)\,dq-i(\mu_{12}^{(0)}+\mu_{12}^{(2)})p_2(t)+
&  \\
\qquad \quad -i(\mu_{13}^{(0)}+\mu_{13}^{(2)})p_3(t)
-i(\nu_{12}^{(0)}+\nu_{12}^{(2)})p_2^\dagger(t)-i(\nu_{13}^{(0)}+%
\nu_{13}^{(2)})p_3^\dagger(t), &  \\
\vspace{1mm} l_2(t)=-i\omega_2 p_2(t)+i\lambda_2\int_{\mathbb{R}%
}B_2(q,t)\,dq+i\tilde\lambda_2\int_{\mathbb{R}}B(q,t)\,dq-i(\mu_{12}^{(0)}+%
\mu_{12}^{(2)})p_1(t)+ &  \\
\qquad \quad -i(\mu_{23}^{(0)}+\mu_{23}^{(2)})p_3(t)
+i(\nu_{12}^{(0)}+\nu_{12}^{(2)})p_1^\dagger(t)-i(\nu_{23}^{(0)}+%
\nu_{23}^{(2)})p_3^\dagger(t), &  \\
\vspace{1mm} l_3(t)=-i\omega_3 p_3(t)+i\lambda_3\int_{\mathbb{R}%
}B_3(q,t)\,dq+i\tilde\lambda_3\int_{\mathbb{R}}B(q,t)\,dq-i(\mu_{13}^{(0)}+%
\mu_{13}^{(2)})p_1(t)+ &  \\
\qquad \quad -i(\mu_{23}^{(0)}+\mu_{23}^{(2)})p_2(t)
+i(\nu_{13}^{(0)}+\nu_{13}^{(2)})p_1^\dagger(t)+i(\nu_{23}^{(0)}+%
\nu_{23}^{(2)})p_2^\dagger(t), \label{26b} &
\end{array}%
\right.
\end{equation}
which are all linear in their entries, and these other functions, which are
not linear:
\begin{equation}
\left\{
\begin{array}{ll}
nl_1(t)=-i(\mu_{12}^{(1)}-\mu_{12}^{(2)})N_3(t)p_2(t)-i(\mu_{13}^{(1)}-%
\mu_{13}^{(2)})N_2(t)p_3(t)+ &  \\
\qquad \quad
-i(\nu_{12}^{(1)}-\nu_{12}^{(2)})N_3(t)p_2^\dagger(t)-i(\nu_{13}^{(1)}-%
\nu_{13}^{(2)})N_2(t)p_3^\dagger(t)+ &  \\
\qquad \quad -i (\mu_{23}^{(1)}-\mu_{23}^{(2)})p_1(t)(p_2^\dagger(t)
p_3(t)+p_3^\dagger(t) p_2(t))+ &  \\
\qquad \quad -i (\nu_{23}^{(1)}-\nu_{23}^{(2)})p_1(t)(p_2^\dagger(t)
p_3^\dagger(t)+p_3(t) p_2(t)), &  \\
\vspace{1mm} nl_2(t)=-i(\mu_{12}^{(1)}-\mu_{12}^{(2)})N_3(t)p_1(t)-i(%
\mu_{23}^{(1)}-\mu_{23}^{(2)})N_1(t)p_3(t)+ &  \\
\qquad \quad
+i(\nu_{12}^{(1)}-\nu_{12}^{(2)})N_3(t)p_1^\dagger(t)-i(\nu_{23}^{(1)}-%
\nu_{23}^{(2)})N_1(t)p_3^\dagger(t)+ &  \\
\qquad \quad -i (\mu_{13}^{(1)}-\mu_{13}^{(2)})p_2(t)(p_1^\dagger(t)
p_3(t)+p_3^\dagger(t) p_1(t))+ &  \\
\qquad \quad -i (\nu_{13}^{(1)}-\nu_{13}^{(2)})p_2(t)(p_1^\dagger(t)
p_3^\dagger(t)+p_3(t) p_1(t)), &  \\
\vspace{1mm} nl_3(t)=-i(\mu_{13}^{(1)}-\mu_{13}^{(2)})N_2(t)p_1(t)-i(%
\mu_{23}^{(1)}-\mu_{23}^{(2)})N_1(t)p_2(t)+ &  \\
\qquad \quad
+i(\nu_{13}^{(1)}-\nu_{13}^{(2)})N_2(t)p_1^\dagger(t)+i(\nu_{23}^{(1)}-%
\nu_{23}^{(2)})N_1(t)p_2^\dagger(t)+ &  \\
\qquad \quad -i (\mu_{12}^{(1)}-\mu_{12}^{(2)})p_3(t)(p_1^\dagger(t)
p_2(t)+p_2^\dagger(t) p_1(t))+ &  \\
\qquad \quad -i (\nu_{12}^{(1)}-\nu_{12}^{(2)})p_3(t)(p_1^\dagger(t)
p_2^\dagger(t)+p_2(t) p_1(t)). \label{26c} &
\end{array}%
\right.
\end{equation}

The last two equations in (\ref{26}) can be rewritten as
\begin{equation*}
B_j(q,t)=B_j(q)e^{-i\Omega_j(q)t}+i\lambda_j\int_0^t
p_j(t_1)e^{-i\Omega_j(q)(t-t_1)}\,dt_1
\end{equation*}
and
\begin{equation*}
B(q,t)=B(q)e^{-i\Omega(q)t}+i\int_0^t \sum_{j=1}^3\tilde\lambda_j
p_j(t_1)e^{-i\Omega(q)(t-t_1)}\,dt_1,
\end{equation*}
which, assuming that $\Omega_j(k)=\Omega_j\, k$ and $\Omega(k)=\Omega\, k$, $%
\Omega,\Omega_j>0$, produce
\begin{equation}
\int_{\mathbb{R}}B_j(q,t)\,dq=\int_{\mathbb{R}}B_j(q)e^{-i\Omega_j q
t}\,dq+i\pi\frac{\lambda_j}{\Omega_j}\,p_j(t),  \label{27}
\end{equation}
and
\begin{equation}
\int_{\mathbb{R}}B(q,t)\,dq=\int_{\mathbb{R}}B(q)e^{-i\Omega q t}\,dq+i\pi%
\frac{\sum_{j=1}^3\tilde\lambda_j\,p_j(t)}{\Omega}.  \label{28}
\end{equation}
Now, long but straightforward computations, allow us to rewrite $l_j(t)$ and
$nl_j(t)$ is a simpler form. In particular we find
\begin{equation}
\left\{
\begin{array}{ll}
l_1(t)=-\tilde\omega_1
p_1(t)-\tilde\gamma_{12}p_2(t)-\tilde\gamma_{13}p_3(t)-i\nu_{12}p_2^%
\dagger(t)-i\nu_{13}p_3^\dagger(t)+\eta_1(t), &  \\
l_2(t)=-\tilde\omega_2
p_2(t)-\tilde\gamma_{12}p_1(t)-\tilde\gamma_{23}p_3(t)+i\nu_{12}p_1^%
\dagger(t)-i\nu_{23}p_3^\dagger(t)+\eta_2(t), &  \\
l_3(t)=-\tilde\omega_3
p_3(t)-\tilde\gamma_{13}p_1(t)-\tilde\gamma_{23}p_2(t)+i\nu_{13}p_1^%
\dagger(t)+i\nu_{23}p_2^\dagger(t)+\eta_3(t), \label{29} &
\end{array}%
\right.
\end{equation}
and
\begin{equation}
\left\{
\begin{array}{ll}
nl_1(t)=-i\delta_{12}^\mu N_3(t) p_2(t)-i\delta_{13}^\mu N_2(t)
p_3(t)-i\delta_{12}^\nu N_3(t) p_2^\dagger(t)-i\delta_{13}^\nu N_2(t)
p_3^\dagger(t)+ &  \\
\qquad \quad -i\delta_{23}^\mu
p_1(t)(p_2^\dagger(t)p_3(t)+p_3^\dagger(t)p_2(t))-i\delta_{23}^\nu
p_1(t)(p_2^\dagger(t)p_3^\dagger(t)+p_3(t)p_2(t)), &  \\
nl_2(t)=-i\delta_{12}^\mu N_3(t) p_1(t)-i\delta_{23}^\mu N_1(t)
p_3(t)+i\delta_{12}^\nu N_3(t) p_1^\dagger(t)-i\delta_{23}^\nu N_1(t)
p_3^\dagger(t)+ &  \\
\qquad \quad -i\delta_{13}^\mu
p_2(t)(p_1^\dagger(t)p_3(t)+p_3^\dagger(t)p_1(t))-i\delta_{13}^\nu
p_2(t)(p_1^\dagger(t)p_3^\dagger(t)+p_3(t)p_1(t)), &  \\
nl_3(t)=-i\delta_{13}^\mu N_2(t) p_1(t)-i\delta_{23}^\mu N_1(t)
p_2(t)+i\delta_{13}^\nu N_2(t) p_1^\dagger(t)+i\delta_{23}^\nu N_1(t)
p_2^\dagger(t)+ &  \\
\qquad \quad -i\delta_{12}^\mu
p_3(t)(p_1^\dagger(t)p_2(t)+p_2^\dagger(t)p_1(t))-i\delta_{12}^\nu
p_3(t)(p_1^\dagger(t)p_2^\dagger(t)+p_2(t)p_1(t)). \label{210} &
\end{array}%
\right.
\end{equation}

Here we have introduced the following simplifying notation:
\begin{equation*}
\tilde\omega_l:=i\omega_l+\pi\left(\frac{\lambda_l^2}{\Omega_l}+\frac{%
\tilde\lambda_l^2}{\Omega}\right), \quad
\tilde\gamma_{k,l}:=i\left(\mu_{k,l}^{(0)}+\mu_{k,l}^{(2)}\right)+\frac{\pi}{%
\Omega}\tilde\lambda_k\tilde\lambda_l,
\end{equation*}
\begin{equation*}
\nu_{kl}=\nu_{kl}^{(0)}+\nu_{kl}^{(2)}, \quad
\delta_{kl}^\mu=\mu_{kl}^{(1)}-\mu_{kl}^{(2)}, \quad
\delta_{kl}^\nu=\nu_{kl}^{(1)}-\nu_{kl}^{(2)},
\end{equation*}
for $k,l=1,2,3$, as well as the operator-valued functions:
\begin{equation*}
\eta_j(t)=i\left(\lambda_j\beta_j(t)+\tilde\lambda_j\beta(t)\right),
\end{equation*}
where
\begin{equation*}
\beta_j(t)=\int_{\mathbb{R}}B_j(q)e^{-i\Omega_j q t}dq, \quad\mbox{ and }%
\quad \beta(t)=\int_{\mathbb{R}}B(q)e^{-i\Omega q t}dq.
\end{equation*}

\vspace{2mm}

\textbf{Remark:--} We notice that these equations return those in \cite{all1}
when we put to zero all the coefficients measuring the non-linearity.
Therefore, in this case, they can be explicitly solved.

\vspace{2mm}

Once we have deduced $p_{j}(t)$, we need to compute the DFs $P_{j}(t)$,
which are defined as follows:
\begin{equation}
P_{j}(t):=\left\langle \hat{P}_{j}(t)\right\rangle =\left\langle
p_{j}^{\dagger }(t)p_{j}(t)\right\rangle ,  \label{add1}
\end{equation}%
$j=1,2,3$. Here $\left\langle .\right\rangle $ is a state over the full
system. These states, \cite{bagbook}, are taken to be suitable tensor
products of vector states for $\mathcal{S}_{\mathcal{P}}$ and states on the
reservoir which obey some standard rules ({please }see below). More
in detail, for each operator of the form $X_{\mathcal{S}}\otimes Y_{\mathcal{%
R}}$, $X_{\mathcal{S}}$ being an operator of $\mathcal{S}_{\mathcal{P}}$ and
$Y_{\mathcal{R}}$ an operator of the reservoir, we put
\begin{equation}
\left\langle X_{\mathcal{S}}\otimes Y_{\mathcal{R}}\right\rangle
:=\left\langle \varphi _{n_{1},n_{2},n_{3}},X_{\mathcal{S}}\varphi
_{n_{1},n_{2},n_{3}}\right\rangle \,\omega _{\mathcal{R}}(Y_{\mathcal{R}}).
\label{add2}
\end{equation}%
Here $\varphi _{n_{1},n_{2},n_{3}}$ is one of the vectors introduced at the
beginning of this section, and each $n_{j}$ represents, as discussed before,
the tendency of $\mathcal{P}_{j}$ to form (or not) some coalition at $t=0$.
Moreover, $\omega _{\mathcal{R}}(.)$ is a state on $\mathcal{R}$ satisfying
the following standard properties, \cite{bagbook}:
\begin{equation}
\omega _{\mathcal{R}}(1\!\!1_{\mathcal{R}})=1,\quad \omega _{\mathcal{R}%
}(B_{j}(k))=\omega _{\mathcal{R}}(B_{j}^{\dagger }(k))=0,\quad \omega _{%
\mathcal{R}}(B_{j}^{\dagger }(k)B_{l}(q))=N_{j}(k)\,\delta _{j,l}\delta
(k-q),  \label{211}
\end{equation}%
as well as
\begin{equation}
\omega _{\mathcal{R}}(B(k))=\omega _{\mathcal{R}}(B^{\dagger }(k))=0,\quad
\omega _{\mathcal{R}}(B^{\dagger }(k)B(q))=N(k)\,\delta (k-q),
\label{211bis}
\end{equation}%
for some suitable functions $N_{j}(k)$ and $N(k)$, which we take here to be
constant in $k$: $N_{j}(k)=N_{j}$ and $N(k)=N$. Also, we assume $\omega _{%
\mathcal{R}}(B_{j}(k)B_{l}(q))=\omega _{\mathcal{R}}(B(k)B(q))=0$, for all $j
$ and $l$. The reason why we use the state in (\ref{add2}) is because it
describes, in our framework, the fact that, at $t=0$, $\mathcal{P}_{j}$'s
decision is $n_{j}$, while the overall feeling of the voters $\mathcal{R}_{j}
$ is $N_{j}$, and that of the undecided ones is $N$. Of course, these might
appear as oversimplifying assumptions, but they {still }produce in
many concrete applications, rather interesting dynamics for the model.

\subsection{The solution}

\label{sectII1}

{To begin with, }we consider now a simple but still non-trivial
situation, which allows us to write the differential equations of the system
in a reasonably simple way and to find an approximate solution. {This
}suggests a strategy which can be easily generalized to other situations.
This is, in fact, what we will do in the last part of this section.

Let us assume for the moment that the coefficients in $\delta h$ are such
\begin{equation*}
\delta _{13}^{\mu }=\delta _{13}^{\nu }=\delta _{23}^{\mu }=\delta
_{23}^{\nu }=\delta _{12}^{\nu }=0,
\end{equation*}%
while $\delta _{12}^{\mu }=\mu _{12}^{(1)}-\mu _{12}^{(2)}\neq 0$, and for
simplicity we call this difference $\delta $: $\delta =\delta _{12}^{\mu }$.
This makes the system non-linear, but not extremely complicated (at least
{not} from the point of view of the notation). The first three
equations of system (\ref{26}), together with their adjoints, can be
rewritten as
\begin{equation}
\dot{P}(t)=TP(t)+\eta (t)+i\delta \Lambda (P(t)),  \label{41}
\end{equation}%
where we have introduced the following vectors:
\begin{equation*}
P(t)=\left(
\begin{array}{c}
p_{1}(t) \\
p_{2}(t) \\
p_{3}(t) \\
p_{1}^{\dagger }(t) \\
p_{2}^{\dagger }(t) \\
p_{3}^{\dagger }(t) \\
\end{array}%
\right) ,\quad \eta (t)=\left(
\begin{array}{c}
\eta _{1}(t) \\
\eta _{2}(t) \\
\eta _{3}(t) \\
\eta _{1}^{\dagger }(t) \\
\eta _{2}^{\dagger }(t) \\
\eta _{3}^{\dagger }(t) \\
\end{array}%
\right) ,\quad \Lambda (P(t))=\left(
\begin{array}{c}
-N_{3}(t)p_{2}(t) \\
-N_{3}(t)p_{1}(t) \\
-p_{3}(t)\left( p_{1}^{\dagger }(t)p_{2}(t)+p_{2}^{\dagger
}(t)p_{1}(t)\right)  \\
p_{2}^{\dagger }(t)N_{3}(t) \\
p_{1}^{\dagger }(t)N_{3}(t) \\
\left( p_{1}^{\dagger }(t)p_{2}(t)+p_{2}^{\dagger }(t)p_{1}(t)\right)
p_{3}^{\dagger }(t) \\
\end{array}%
\right) ,
\end{equation*}%
as well as the matrix
\begin{equation*}
T=\left(
\begin{array}{cccccc}
-\tilde{\omega}_{1} & -\tilde{\gamma}_{12} & -\tilde{\gamma}_{13} & 0 &
-i\nu _{12} & -i\nu _{13} \\
-\tilde{\gamma}_{12} & -\tilde{\omega}_{2} & -\tilde{\gamma}_{23} & i\nu
_{12} & 0 & i\nu _{23} \\
-\tilde{\gamma}_{13} & -\tilde{\gamma}_{23} & -\tilde{\omega}_{3} & i\nu
_{13} & i\nu _{23} & 0 \\
0 & i\nu _{12} & i\nu _{13} & -\overline{\tilde{\omega}_{1}} & -\overline{%
\tilde{\gamma}_{12}} & -\overline{\tilde{\gamma}_{13}} \\
-i\nu _{12} & 0 & i\nu _{23} & -\overline{\tilde{\gamma}_{12}} & -\overline{%
\tilde{\omega}_{2}} & -\overline{\tilde{\gamma}_{23}} \\
-i\nu _{13} & -i\nu _{23} & 0 & -\overline{\tilde{\gamma}_{13}} & -\overline{%
\tilde{\gamma}_{23}} & -\overline{\tilde{\omega}_{3}} \\
&  &  &  &  &
\end{array}%
\right) .
\end{equation*}%
Solving exactly equation (\ref{41}) is quite hard, if not impossible, due to
the non-linearity included in $\Lambda (P(t))$. However, it is easy to set
up a recursive approximation approach which might converge to, or at least
approximate, the solution. The idea is simple, and {it }works better
under the assumption that $\delta $ is sufficiently small. In this case we
replace (\ref{41}) with the following, much simpler, equation: $\dot{P}%
_{0}(t)=TP_{0}(t)+\eta (t)$, which is linear and can be easily solved. The
solution is
\begin{equation*}
P_{0}(t)=e^{Tt}\left( P(0)+\int_{0}^{t}e^{-Tt_{1}}\eta
_{0}(t_{1})dt_{1}\right) ,
\end{equation*}%
where we have introduced, for reasons which will be clear in a moment, $\eta
_{0}(t)\equiv \eta (t)$. We can now use this {zero-th order approximation}
of $P(t)$ in $\Lambda (P(t))$, in equation (\ref{41}), which becomes $\dot{P}%
_{1}(t)=TP_{1}(t)+\eta _{1}(t)$, where $\eta _{1}(t)=\eta _{0}(t)+i\delta
\Lambda (P_{0}(t))$. Notice that $\eta _{1}(t)$ is now a known function. The
solution of this equation is
\begin{equation*}
P_{1}(t)=e^{Tt}\left( P(0)+\int_{0}^{t}e^{-Tt_{1}}\eta
_{1}(t_{1})dt_{1}\right) .
\end{equation*}%
Of course, we can iterate the procedure, and the $n$-th approximation is
\begin{equation}
P_{n}(t)=e^{Tt}\left( P(0)+\int_{0}^{t}e^{-Tt_{1}}\eta
_{n}(t_{1})dt_{1}\right) ,  \label{42}
\end{equation}%
where $\eta _{n}(t)=\eta (t)+i\delta \Lambda (P_{n-1}(t))$, for $n\geq 1$.
Hence, at least in principle, we can reach the level of approximation we
want. However, we should also say that it is not guaranteed that the
sequence $\{P_{n}(t)\}$ really converges to the solution of (\ref{41}), even
if this might appear rather reasonable. Similar problems often occur when
non-linear differential equations are considered, as it happens in our
system. Summarizing, we cannot, a priori, say that (i) $\lim_{n\rightarrow
\infty }P_{n}(t)$ exists (in some suitable topology), and (ii) even if it
exists, if this limit is the solution of equation (\ref{41}). Nevertheless,
what we can safely say, is that $P_{n}(t)$ is a certain approximation of $%
P(t)$, and we suspect that this approximation is sufficiently good for small
$\delta $ and $t$, and for large $n$. Of course, more could be said only after
numerical computations or looking for some a priori estimates. This is indeed part of our
work in progress.

However, there is a situation in which the computations can be carried out explicitly. In fact, if $\delta
_{kl}^{\mu }=\delta _{kl}^{\nu }=0$ for all $k,l$, then, as already
observed, the equations reduce to those for the linear system\footnote{Incidentally we observe that this does not imply that all the parameters of $\delta h$ are zero. It only means that they coincide in pairs.}.  Hence, they are exactly solvable and the
result has been discussed in \cite{all1}. Looking at the analytical form of $%
\delta h$ in (\ref{24b}), this can be understood since {it} correspond%
{s} to the fact that, for instance, $\mathcal{P}_{1}$ and $\mathcal{P}%
_{2}$ react with the same strength to the will of $\mathcal{P}_{3}$ to
{either }create or not an alliance.

In the next section we will briefly show that we can consider cases other than the one considered above.  In fact, a general solution can also be found even when the parameters in $\delta h$ are different from each other.

\subsection{A more general situation}

It is clear that when we give up the working assumptions we have considered
above (i.e. $\delta _{13}^{\mu }=\delta _{13}^{\nu }=\delta _{23}^{\mu
}=\delta _{23}^{\nu }=\delta _{12}^{\nu }=0$), the explicit form of the non-linear term $i\delta \Lambda
(P(t))$ changes. This is due to the presence of several parameters and not of just one. Consequently, it is convenient to modify the
strategy and this can be done as follows: the starting point is the equation $$\dot{P}(t)=TP(t)+\eta (t)+\tilde{\Lambda}(P(t)),$$ where $\tilde{\Lambda}(P(t))$ is
the strong non-linear contribution which extends the term $i\delta \Lambda (P(t))$ in (\ref{41}). Introducing now $P_1(t)=e^{-Tt}P(t)$, $\eta_1(t)=e^{-Tt}\eta (t)$ and $\tilde{\Lambda}_1(P_1(t))=e^{-Tt}\tilde{\Lambda}(e^{Tt}P(t))$, the equation for $P_1(t)$ becomes $$\dot{P_1}(t)=\eta_1 (t)+\tilde{\Lambda}_1(P_1(t)),$$ which can be still be re-written in a more convenient form by introducing further the $\eta_2(t)=\int_0^t\eta_1(t_1)dt_1$, and the new unknown $P_2(t)=P_1(t)-\eta_2(t)$. In fact, calling now $\tilde{\Lambda}_2(P_2(t))=\tilde{\Lambda}_1(P_2(t)+\eta_2(t))$, we get a very simple differential equation,
$$
\dot P_2(t)=\tilde{\Lambda}_2(P_2(t)),
$$
whose formal solution is \be\int dP_2 \tilde{\Lambda}_2^{-1}(P_2)=t+\alpha,\label{43}\en $\alpha$ being an integration constant. Of course, this solution is {\em formal} because of several reasons: firstly, we don't know a priori if $\tilde{\Lambda}_2^{-1}(P_2)$ exists. Secondly, we are not sure we can compute its integral. Thirdly, we are working with operators (and not with simple functions). This makes the situation even more complicated. However, in principle, formula (\ref{43}) produces the solution of the general problem, without any approximation. Hence, from a certain point of view, it looks much more interesting than the solution deduced in the previous section. We will devote a future analysis to a deeper, and more explicit, analysis of the results arising from equation (\ref{43}).

\section{Dynamics of buying and selling}

We have already {remarked }in several papers ({please }see in
particular \cite{baghav1,baghav2} for recent results) that the above
extended hamiltonian framework c{ould} be applied to economics and
finance. We show now that this is true {and in so doing we }chang%
{e} the interpretation of the model considered here. In particular,
we will now discuss that the resulting framework becomes akin to a formal
structure which can describe the dynamics of buying and selling ({of
financial assets for instance)}. However, the framework does not explicitly
provide for a mechanism by which prices can be generated. We note first that
when considering the different terms which are part of $H=h+\delta h$, we
can in effect make an argument that the hamiltonians $H_{0}$, $H_{PBs}$, $%
H_{PB}$ are associated with public information which occurs at a macroscale,
since they are connected with some reservoirs which describe in fact (%
{please }see below), large groups of people. As is reported in \cite%
{MS2000}, the reaction of traders on this public information is then
transferred onto smaller scales, i.e. to traders themselves. The scale at
which this happens is cast by the hamiltonians $H_{int,l}$ and $h$.

The above framework, we insist, hints back to {the binary choice of
either }buying and selling. The key reason for that is that the eigenvalues
of the number operators are either `$0$' or `$1$'. The financial system
which we want to emulate with $H=h+\delta h$ must contain interactions and
hence we can not be satisfied with just using $H_{0}$. This interaction in
the framework proposed here can be either at the macroscale and/or the
microscale (i.e. between the traders). The division of two grand types of
information, i.e. public and private information occurs typically (and
intuitively) at respectively the macroscale and the microscale. One can of
course be rigorous about this. Work by \cite{Scheinkman2004} for instance
shows that private information has no effect at all on traders when they
behave in a rational expectations model.

To make sure we use some reference framework from the economics literature
on how to properly define public versus private information, we resort to
\cite{BHR2013} who define public information {as} having the
potential to be known by everyone, whilst private information may be known
by one single individual. In our situation, the decision functions $P_{j}(t)$
describe the will of the three traders\footnote{%
Of course, we are sticking here to just three traders because of our
previous application to political alliances, but it is not difficult to
extend the model to more traders.}, $\mathcal{P}_{1}$, $\mathcal{P}_{2}$ and
$\mathcal{P}_{3}$, to buy (zero) or to sell (one) some assets. This choice
is driven by public information (i.e. by $\mathcal{R}_{j}$ and $\mathcal{R}%
_{und}$, see below) and by private information (i.e. by the mutual
interaction between the traders).

On the basis of public information, traders can adjust their portfolio
holdings and this, as \cite{BHR2013} indicates, can affect prices in the
market. The opposite may well be true in the case of private information,
where a single party profits but with no {necessary }effect on price
behavior. What is interesting is the statement by \cite{BHR2013} that almost
always (see p. 224), will there be processes operating which will
`publicize' private information. Please consider again $H_{PBs}$, $H_{PB}$
which was mentioned in the context of the politics example above. Assume we
have three traders who have the binary elemental task of either selling or
buying. Denote $H=h+\delta h$ as the hamiltonian which describes the
dynamics of buying and selling over time, under the influence of both
private and public information. Besides the no-interaction hamiltonian, $%
H_{0}$, the dynamic drivers which are associated to public information are,
as stated, $H_{PBs}$ and $H_{PB}$. From an economics point of view, the
baths $\mathcal{R}_{1}$, $\mathcal{R}_{2}$ and $\mathcal{R}_{3}$ now signify
a vast collection of \textit{informed} traders with which our three traders
interact with (in view of performing the elemental task of buying and
selling). Whilst the bath $\mathcal{R}_{und}$ consists of a vast collection
of traders, who can be interpreted as \textit{noise }traders. This can be
easily achieved in our model by assuming some randomness of the $\tilde{%
\lambda}_{j}$ in (\ref{22}). A question arises whether we can be rigorous in
defining those two types of traders. In \cite{HK2013} noise traders are
defined as traders who act upon information which is more often than not,
spurious information. Informed traders have at their command information
which can be objectively used in decisions involving buying or selling.

The contribution of $H_{PBs}$ in $H$ (i.e. the {full} driver of the
dynamics of buying and selling) describes the interaction between the three
traders and the baths of informed traders. Clearly, we want to point out
that this interaction is occurring via the medium of public information,
given the size of the baths. The mechanism that $p_{j}B_{j}^{\dagger }(k)$
describes now leads to (say) the action of selling by traders given {%
that }some public information (from informed traders) has been released that
`selling' is what one should do. In identical fashion do we argue for a
buying signal when $B_{j}(k)p_{j}^{\dagger }$ occurs. But note also the
contribution of $H_{PB}$, which now influences traders to sell or buy given
public information coming from noise traders. Both those buying and selling
signals, whether they either derive from the interaction with the baths $%
\mathcal{R}_{1}$, $\mathcal{R}_{2}$ and $\mathcal{R}_{3}$ or the $\mathcal{R}%
_{und}$ bath, have the potential to ultimately influence price setting given
that public information is at stake.

What is contained in $H_{int^{\prime }l\text{ }}$ and $\delta h$ are
communications between traders, without recourse to the public information
baths. We have three traders, and by virtue of this very small size, it is
perfectly intuitive to call the information, upon which traders make
decisions within this interaction setting, to be private information. But as
has been remarked above, whilst in $H_{int^{\prime }l\text{ }}$ the
individual's traders decision of buying and selling does not affect their
`partner' traders, there is a very explicit dependence between the
individual's traders built in when considering $\delta h$. However, private
information as such is not expected to influence price behavior. Private
information, as we have remarked above, seems to be subject to the act of
`publicizing' private information. Well known notions like information
leakage and uncertainty creation can be following from such an act. See \cite%
{BHR2013} and \cite{MA2011} for a discussion. In \cite{MA2011} (see also
\cite{HK2013}), information leakage is defined as \textquotedblleft
situations where agents wish to reveal truthfully their private possessed
information to others\textquotedblright . Such type of release of
information invites in cooperation amongst agents and it also very clearly
creates an interdependence between agents. Information leakage can be
selective, i.e. agent 1 can release information only to agent 2 and thereby
alienate agent 3. Similarly, in the case of so called `uncertainty
creation', information is created which is on purpose false or erroneous, so
as to induce other agents in error so it can serve one's own investment
strategy. This is again an example of private information which, on purpose,
creates dependencies between traders. One can even get more precise by
considering the quality of the private information. Trader 1 can release
private information with noise to trader 2 but without any noise to trader
3. See \cite{BRUNNER2001} (p. 71). One can even be more refined and
introduce so called knowledge operators in the modeling of information. See
again \cite{BRUNNER2001} (p. 4-). Of course, these several different effects
all suggest the relevance of the full hamiltonian $H$ in (\ref{24b}), and
importance {is given }to its various contributions also in this
economics context. Incidentally, this means that the differential equations
governing this {particular} application are again (\ref{26}), so that
the same approximation procedure discussed in Section \ref{sectII1} can be
adopted. Needless to say, that for this particular application, our next
step will surely be to produce numerical solutions and/or analytical
estimates. This is, for the present model, a hard task. However, it can be
easily done in the linear case, simply {by }adapting what we have
done in \cite{all1} to the present situation. Before doing that, we would
like to mention that this analogy presented here in this section does query
however, how departures from equilibrium can be caused by $H_{int,l\text{ }}$
and $\delta h$ if we align those hamiltonians with the existence of private
information. As such buying and selling ensuing from private information is
unlikely to affect price behavior. Hence, this is unlikely to affect the
equilibrium price obtained out of public information-based buying and
selling.

\subsection{Back to the linear case}

\begin{figure}[ht]
\begin{center}
\includegraphics[width=0.4\textwidth]{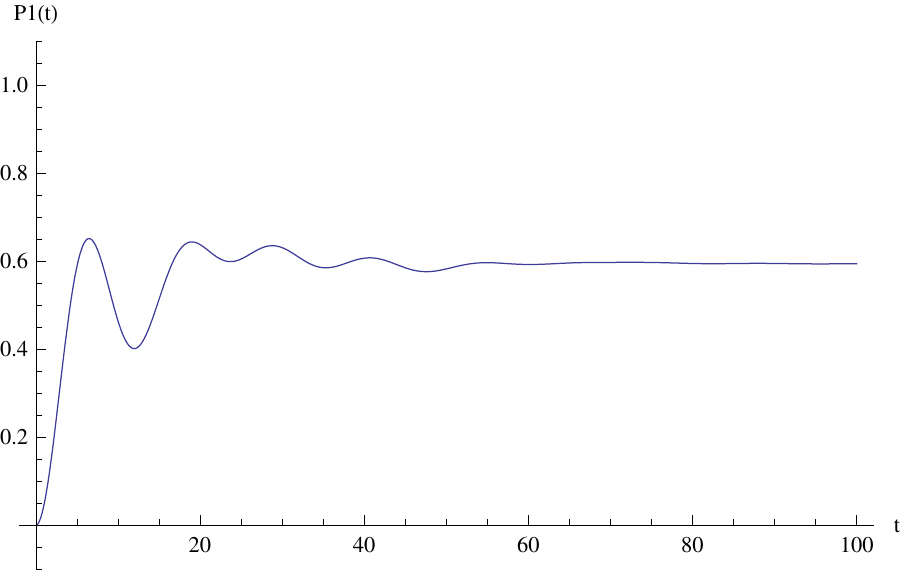}\hspace{8mm} %
\includegraphics[width=0.4\textwidth]{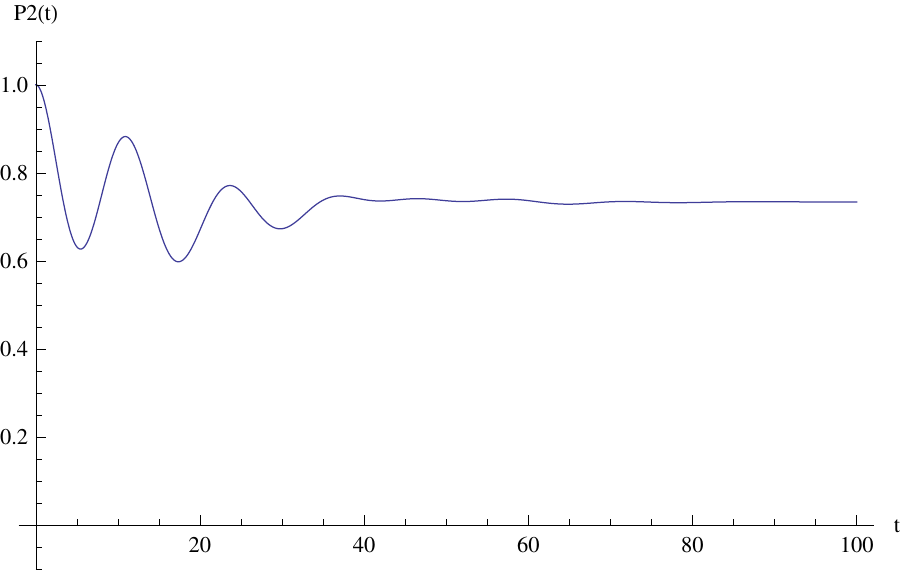}\hfill\\[0pt]
\includegraphics[width=0.4\textwidth]{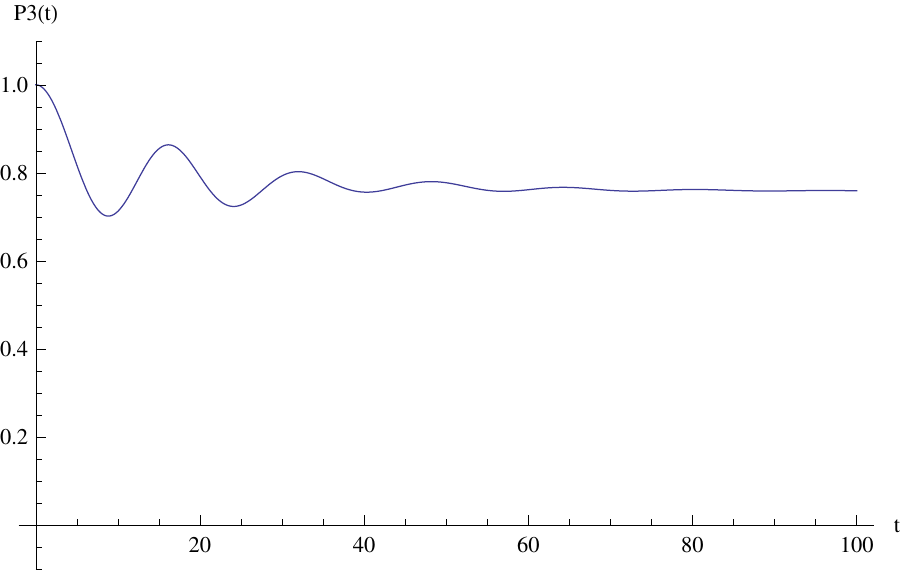}
\end{center}
\caption{{\protect\footnotesize $P_1(t)$ (top left), $P_2(t)$ (top right)
and $P_3(t)$ (bottom) for $\protect\mu_{1,2}^{(0)}=0.2$, $\protect\mu%
_{1,3}^{(0)}=0.1$, $\protect\mu_{2,3}^{(0)}=0.15$, $\protect\nu%
_{k,l}^{(0)}=\tilde\protect\lambda_j=0$, $\protect\omega_1=0.1$, $\protect%
\omega_2=\protect\omega_3=0.2$, $\Omega_1=\Omega_3=1$, $\Omega_2=2$, $%
\Omega=0.1$, $\protect\lambda_1=0.1$, $\protect\lambda_2=0.2$, $\protect%
\lambda_3=0.05$, and $n_1=0$, $n_2=n_3=1$, $N_1=0$, $N_2=N_3=N=1$.}}
\label{fig_a}
\end{figure}

\begin{figure}[ht]
\begin{center}
\includegraphics[width=0.4\textwidth]{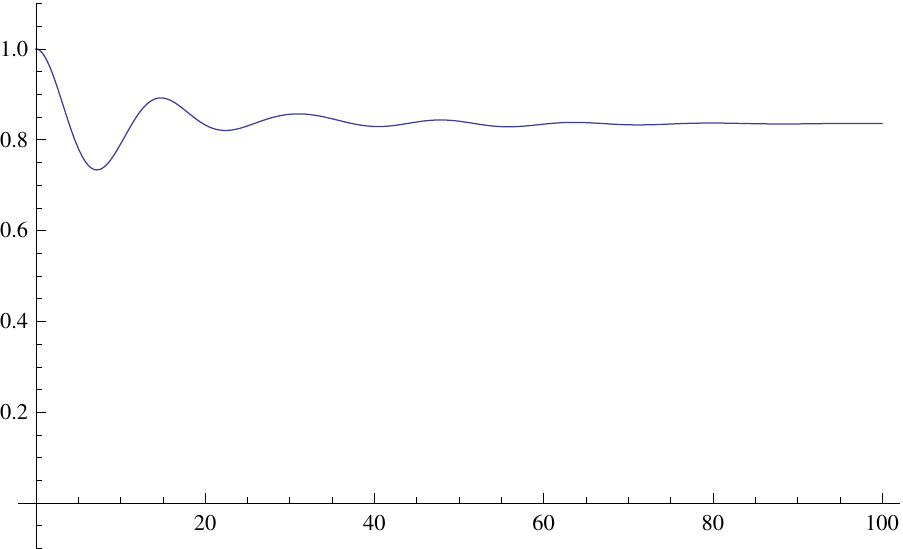}\hspace{8mm} %
\includegraphics[width=0.4\textwidth]{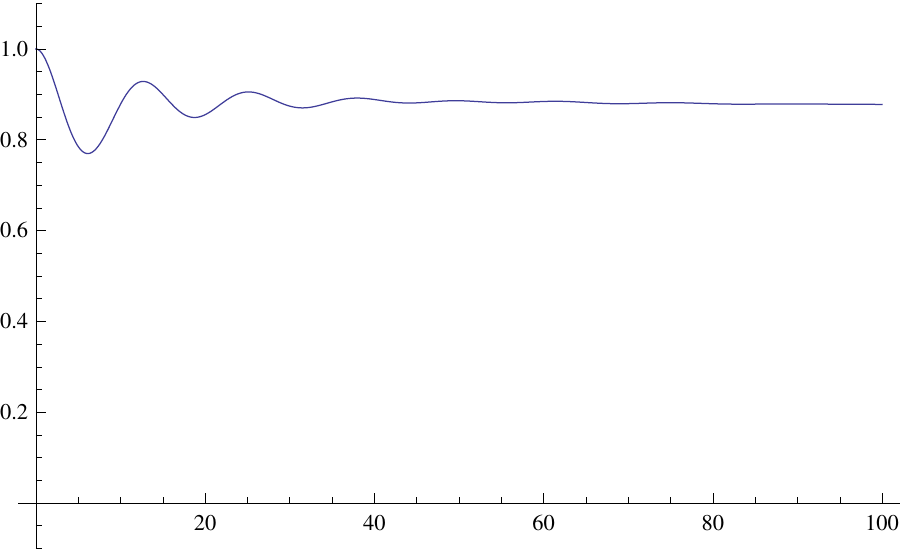}\hfill\\[0pt]
\includegraphics[width=0.4\textwidth]{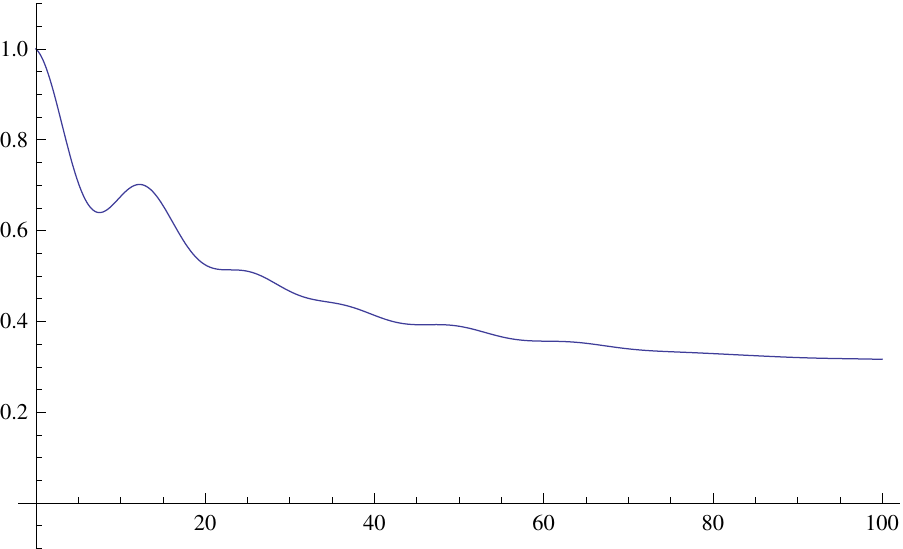}
\end{center}
\caption{{\protect\footnotesize $P_1(t)$ (top left), $P_2(t)$ (top right)
and $P_3(t)$ (bottom) for $\protect\nu_{1,2}^{(0)}=0.1$, $\protect\nu%
_{1,3}^{(0)}=0.08$, $\protect\nu_{2,3}^{(0)}=0.1$, $\protect\mu%
_{k,l}^{(0)}=\tilde\protect\lambda_j=0$, $\protect\omega_1=0.1$, $\protect%
\omega_2=\protect\omega_3=0.2$, $\Omega_1=\Omega_3=1$, $\Omega_2=2$, $%
\Omega=0.1$, $\protect\lambda_1=0.1$, $\protect\lambda_2=0.2$, $\protect%
\lambda_3=0.05$, and $n_1=n_2=n_3=1$, $N_1=N_2=1$, $N_3=N=0$.}}
\label{fig_b}
\end{figure}

In this section we see what happens when $\delta h=0$, i.e. when $\mu
_{kl}^{(1)}=\mu _{kl}^{(2)}=\nu _{kl}^{(1)}=\nu _{kl}^{(2)}=0$ for all $k$
and $l$. In this case, $H=h$, which is quadratic in creation and
annihilation operators, and the differential equations (\ref{41}) become
linear. Essentially, we go back to what we have done, in a political
context, in \cite{all1}. In fact the numerical plots are completely
analogous. For instance, Figures \ref{fig_a} and \ref{fig_b} show the three
DFs for two different choices of the parameters of the hamiltonian and for
certain initial conditions ({please }see {the} figure's
caption). These two sets of parameters correspond to two different
situations. In the first {situation}, Figure \ref{fig_a}, each trader
interacts with its related $\mathcal{R}_{j}$, but not with $\mathcal{R}_{und}
$. They also interact amongst them, but only adopt the mutual different
mechanism described by terms like $p_{1}^{\dagger }p_{2}+p_{2}^{\dagger
}p_{1}$ in (\ref{22}). In {the second situation,} Figure \ref{fig_b},
we describe a similar situation but with the difference that the only
possible interaction between the traders is of {the }cooperative
type: only terms like $p_{1}^{\dagger }p_{2}^{\dagger }+p_{2}p_{1}$ survive.

From both figures we see that, with these choices of parameters and initial
conditions, the three DFs begin oscillating and then reach some asymptotic
value, which is not just zero or one. In \cite{all1} we have discussed why
this is so, and when a sharp result can be really deduced. The conclusion,
here, is that it is quite unlikely that the traders reach some decision they
are completely satisfied {with}. However, see for instance $P_{1}(t)$
and $P_{2}(t)$ {in} Figure \ref{fig_b}, the asymptotic values of both
these DFs are close to one. {Hence,} we see that the decision process
produce{s} a sort of unique decision. On the other hand, $\mathcal{P}%
_{3}$ is not really sure of what he has to do, since $P_{3}(t)$ for large $t$
approaches 0.4, which is not so close to zero.

\vspace{2mm}

\begin{figure}[ht]
\begin{center}
\includegraphics[width=0.4\textwidth]{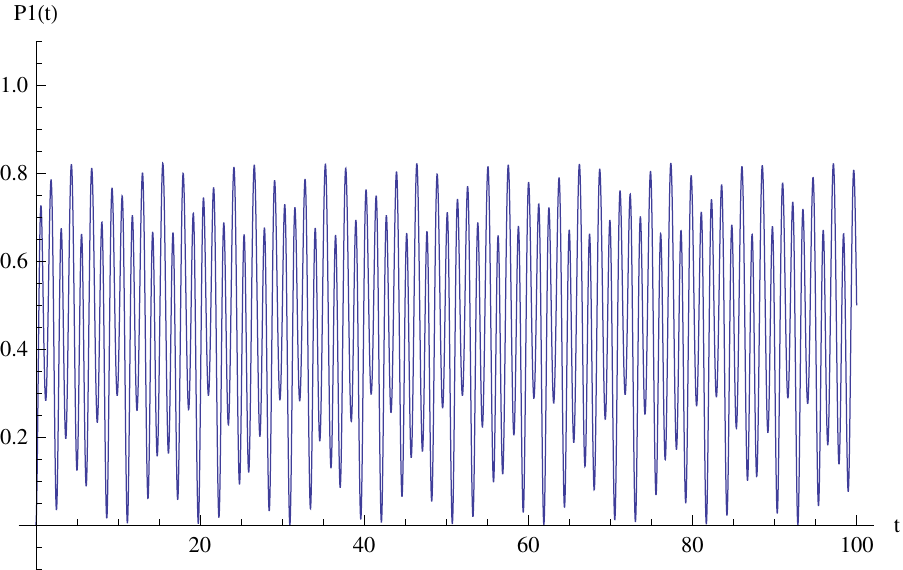}\hspace{8mm} %
\includegraphics[width=0.4\textwidth]{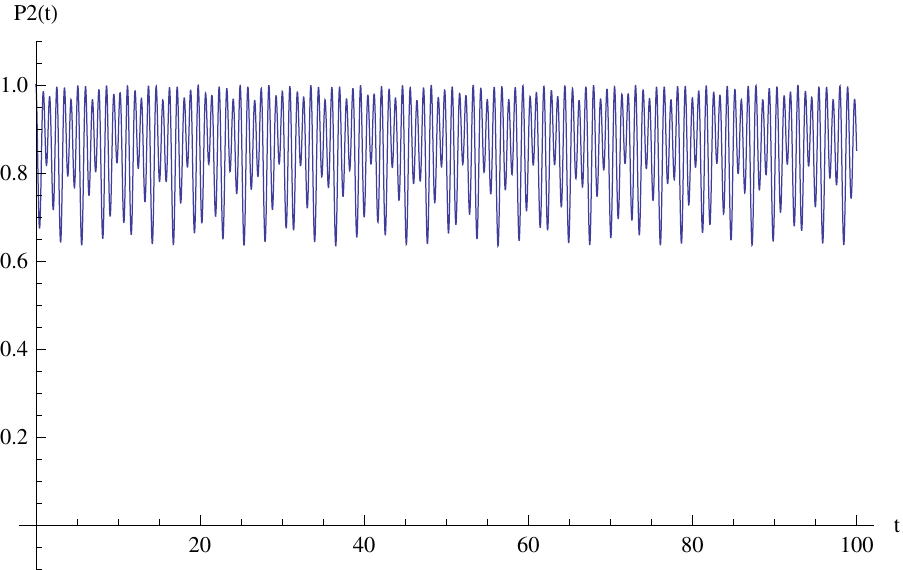}\hfill\\[0pt]
\includegraphics[width=0.4\textwidth]{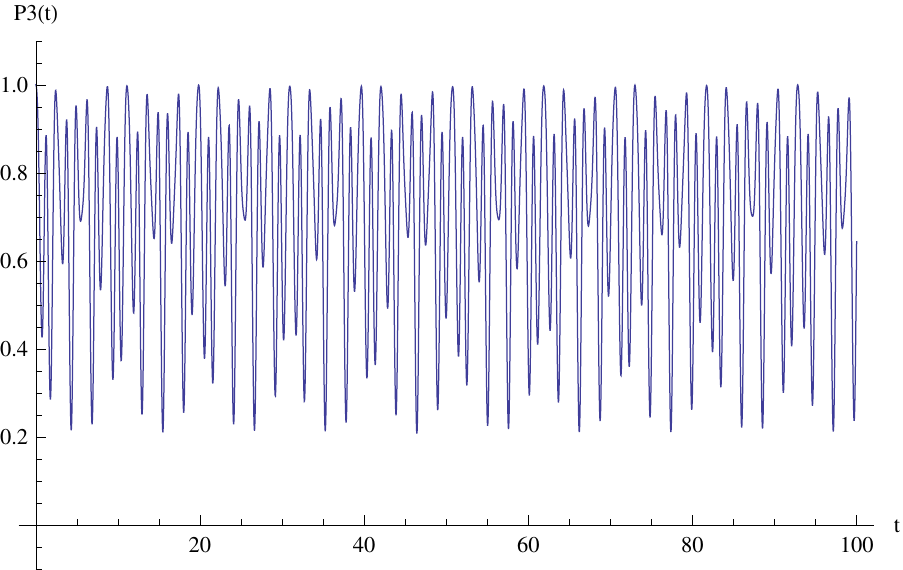}
\end{center}
\caption{{\protect\footnotesize $P_1(t)$ (top left), $P_2(t)$ (top right)
and $P_3(t)$ (bottom) for $\protect\nu_{1,2}^{(0)}=0.1$, $\protect\nu%
_{1,3}^{(0)}=0.08$, $\protect\nu_{2,3}^{(0)}=0.1$, $\protect\mu%
_{1,2}^{(0)}=2 $, $\protect\mu_{1,3}^{(0)}=1$, $\protect\mu_{2,3}^{(0)}=3$, $%
\tilde\protect\lambda_j=\protect\lambda_j=0$, $\protect\omega_1=0.1$, $%
\protect\omega_2=\protect\omega_3=0.2$, $\Omega_1=\Omega_3=\Omega=0.1$, $%
\Omega_2=0.2$ and $n_1=0$, $n_2=n_3=1$, $N_1=N_2=1$, $N_3=N=0$.}}
\label{fig_c}
\end{figure}

A different story is described by Figure \ref{fig_c}, where we are assuming
that the traders only interact among themselves and not with any $\mathcal{R}%
_{j}$ or with $\mathcal{R}_{und}$. When this happens {it }is clear
that none of the traders is able to reach a final decision on whether to buy
or sell the asset. They just oscillate between different \emph{feelings},
but a conclusion can only be reached when the traders also have some input
from the larger sets of informed and noise traders.

\section{Conclusions}

In this paper we have shown how to use operatorial techniques, and an
Heisenberg-like dynamics, to describe two different, but somehow related,
decision making processes. One {such process is }related to political
alliances and {the other process relates }to buy and sell phenomena.
A non-linear model  which extends th{e model} proposed in \cite{all1}%
, has been introduced and an approximate procedure for the solution of the
related equations of motion has also been proposed. We postpone to a second
part of the paper the explicit analysis of these solutions, and a detailed
analysis of the role of the parameters of the model. We claim that, for
small values of the parameters governing the non-linearity, and for time
intervals sufficiently small, these solutions do not differ significantly
from those deduced in \cite{all1}. It is of course of interest to check what
happens for longer intervals, and this {will form} part of a {%
forthcoming} project. Also, it can be interesting to extend the system
described in Figure \ref{figscheme} adding more arrows. In particular, a
natural extension of the model discussed in Section \ref{sectII} can be
constructed by admitting that, for instance, $\mathcal{P}_{1}$ also
interacts with $\mathcal{R}_{2}$ and $\mathcal{R}_{3}$ (i.e. to try to
convince them to change their intentions of vote).

\section*{Acknowledgements}

This work was partially supported by the University of Palermo and by
G.N.F.M. The authors thank Prof. Andrei Khrennikov for many useful
discussions. F.B. acknowledges the warm hospitality of the IQSCS institute
at the University of Leicester.

\renewcommand{\theequation}{A.\arabic{equation}}

\section*{Appendix: A few results on the number representation}

To keep the paper self-contained, we discuss here {a }few important
facts in quantum mechanics and in the so--called number representation.

Let $\mathcal{H}$ be a Hilbert space, and $B(\mathcal{H})$ the set of all
the (bounded) operators on $\mathcal{H}$. Let $\mathcal{S}$ be our physical
system, and ${\mathfrak{A}}$ the set of all the operators useful for a
complete description of $\mathcal{S}$, which includes the \emph{observables}
of $\mathcal{S}$. For simplicity, it is convenient (but not really
necessary) to assume that ${\mathfrak{A}}$ coincides with $B(\mathcal{H})$
itself. The description of the time evolution of $\mathcal{S}$ is related to
a self--adjoint operator $H=H^{\dagger }$ which is called the \emph{%
hamiltonian} of $\mathcal{S}$, and which in standard quantum mechanics
represents the energy of $\mathcal{S}$. In this paper, we have adopted the
so--called \emph{Heisenberg} representation, in which the time evolution of
an observable $X\in {\mathfrak{A}}$ is given by
\begin{equation}
X(t)=\exp (iHt)X\exp (-iHt),  \label{a1}
\end{equation}%
or, equivalently, by the solution of the differential equation
\begin{equation}
\frac{dX(t)}{dt}=i\exp (iHt)[H,X]\exp (-iHt)=i[H,X(t)],  \label{a2}
\end{equation}%
where $[A,B]:=AB-BA$ is the \emph{commutator} between $A$ and $B$. The time
evolution defined in this way is a one--parameter group of automorphisms of $%
{\mathfrak{A}}$.

An operator $Z\in{\mathfrak{A}}$ is a \emph{constant of motion} if it
commutes with $H$. Indeed, in this case, equation (\ref{a2}) implies that $%
\dot Z(t)=0$, so that $Z(t)=Z$ for all $t$.

In some previous applications, \cite{bagbook}, a special role was played by
the so--called \emph{canonical commutation relations}. Here, these are
replaced by the so--called \emph{canonical anti--commutation relations}
(CAR): we say that a set of operators $\{a_{\ell },\,a_{\ell }^{\dagger
},\ell =1,2,\ldots ,L\}$ satisfy the CAR if the conditions
\begin{equation}
\{a_{\ell },a_{n}^{\dagger }\}=\delta _{\ell n}1\!\!1,\hspace{8mm}\{a_{\ell
},a_{n}\}=\{a_{\ell }^{\dagger },a_{n}^{\dagger }\}=0  \label{a3}
\end{equation}%
hold true for all $\ell ,n=1,2,\ldots ,L$. Here, $1\!\!1$ is the identity
operator and $\{x,y\}:=xy+yx$ is the \emph{anticommutator} of $x$ and $y$.
These operators, which are widely analyzed in any {quantum mechanics }%
textbook (see, for instance, \cite{mer,rom}) are those which are used to
describe $L$ different \emph{modes} of fermions. From these operators we can
construct $\hat{n}_{\ell }=a_{\ell }^{\dagger }a_{\ell }$ and $\hat{N}%
=\sum_{\ell =1}^{L}\hat{n}_{\ell }$, which are both self--adjoint. In
particular, $\hat{n}_{\ell }$ is the \emph{number operator} for the $\ell $%
--th mode, while $\hat{N}$ is the \emph{number operator of $\mathcal{S}$}.
Compared with bosonic operators, the operators introduced here satisfy a
very important feature: if we try to square them (or to rise to higher
powers), we simply get zero: for instance, from (\ref{a3}), we have $a_{\ell
}^{2}=0$. This is related to the fact that fermions satisfy the Fermi
exclusion principle \cite{rom}.

The Hilbert space of our system is constructed as follows: we introduce the
\emph{vacuum} of the theory, that is a vector $\varphi_{\mathbf{0}}$ which
is annihilated by all the operators $a_\ell$: $a_\ell\varphi_{\mathbf{0}}=0$
for all $\ell=1,2,\ldots,L$. Such a non zero vector surely exists. Then we
act on $\varphi_{\mathbf{0}}$ with the operators $a_\ell^\dagger$ (but not
with higher powers, since these powers are simply zero!):
\begin{equation}
\varphi_{n_1,n_2,\ldots,n_L}:=(a_1^\dagger)^{n_1}(a_2^\dagger)^{n_2}\cdots
(a_L^\dagger)^{n_L}\varphi_{\mathbf{0}},  \label{a4}
\end{equation}
$n_\ell=0,1$ for all $\ell$. These vectors form an orthonormal set and are
eigenstates of both $\hat n_\ell$ and $\hat N$: $\hat
n_\ell\varphi_{n_1,n_2,\ldots,n_L}=n_\ell\varphi_{n_1,n_2,\ldots,n_L}$ and $%
\hat N\varphi_{n_1,n_2,\ldots,n_L}=N\varphi_{n_1,n_2,\ldots,n_L},$ where $%
N=\sum_{\ell=1}^Ln_\ell$. Moreover, using the CAR, we deduce that
\begin{equation*}
\hat
n_\ell\left(a_\ell\varphi_{n_1,n_2,\ldots,n_L}\right)=(n_\ell-1)(a_\ell%
\varphi_{n_1,n_2,\ldots,n_L})
\end{equation*}
and
\begin{equation*}
\hat
n_\ell\left(a_\ell^\dagger\varphi_{n_1,n_2,\ldots,n_L}\right)=(n_%
\ell+1)(a_l^\dagger\varphi_{n_1,n_2,\ldots,n_L}),
\end{equation*}
for all $\ell$. Then $a_\ell$ and $a_\ell^\dagger$ are called the \emph{%
annihilation} and the \emph{creation} operators. Notice that, in some sense,
$a_\ell^\dagger$ is {also} an annihilation operator since, acting on
a state with $n_\ell=1$, we destroy that state.

The Hilbert space $\mathcal{H}$ is obtained by taking the linear span of all
these vectors. Of course, $\mathcal{H}$ has a finite dimension. In
particular, for just one mode of fermions, $dim(\mathcal{H})=2$. This also
implies that, contrarily to what happens for bosons, all the fermionic
operators are bounded.

The vector $\varphi_{n_1,n_2,\ldots,n_L}$ in (\ref{a4}) defines a \emph{%
vector (or number) state } over the algebra ${\mathfrak{A}}$ as
\begin{equation}
\omega_{n_1,n_2,\ldots,n_L}(X)=
\langle\varphi_{n_1,n_2,\ldots,n_L},X\varphi_{n_1,n_2,\ldots,n_L}\rangle,
\label{a5}
\end{equation}
where $\langle\,,\,\rangle$ is the scalar product in $\mathcal{H}$. As we
have discussed in \cite{bagbook}, these states are useful to \emph{project}
from quantum to classical dynamics and to fix the initial conditions of the
considered system.

\end{document}